\documentclass{appolb}
\usepackage{amsfonts}                   
\usepackage{graphicx,color}		

\begin{document}
\newtheorem{theorem}{{\sc Theorem}}[section]
\newtheorem{lemma}{{\sc Lemma}}[section]
\newtheorem{corollary}{{\sc Corollary}}[section]
\newtheorem{proposition}{{\sc Proposition}}[section]
\newtheorem{remark}{{\sc Remark}}[section]
\newtheorem{definition}{{\sc Definition}}[section]

\renewcommand*{\thetheorem}{\thesection.\arabic{theorem}.}
\renewcommand*{\thelemma}{\thesection.\arabic{lemma}.}
\renewcommand*{\thecorollary}{\thesection.\arabic{corollary}.}
\renewcommand*{\theproposition}{\thesection.\arabic{proposition}.}
\renewcommand*{\theremark}{\thesection.\arabic{remark}.}
\renewcommand*{\thedefinition}{\thesection.\arabic{definition}.}

\title{The Asymptotic Dependence of Elliptic Random Variables~\thanks{Presented at FENS 2006}}%
\author{Krystyna Jaworska 
\address{Institute of Mathematics and Cryptology, Military University of Technology\\
ul. Kaliskiego 2, 00-908 Warszawa, Poland}}
\maketitle

\begin{abstract}
In this paper, we try to answer the question, whether for bivariate elliptic random variable $X=(X_1,X_2)$
the marginal random variables $X_1$ and $X_2$ are asymptotically dependent.
We show, that for some special form of the characteristic generator of $X$ the answer is positive.\\ 

\noindent
{\bf keywords:}
dependence of extreme events, risk management,

\end{abstract}

\section{Motivation}
In order to give an answer to the question, "What is the origin of the interest of the asymptotic dependence
of elliptic random variables?" one has to go back several dozen years.\\
Already in the years 1950's and 1960's researchers discovered the non-normal behaviour of financial market
data. In the early 1990's an understanding of the methodology underlying financial or insurance extremes
became very important. Traditional statistics mostly concerns the laws governing averages. But when we look
at the largest ( respectively the smallest) elements in a sample, the assumption of normality seems not to 
be reasonable in the number of applications, particularly in finance and insurance. And heavy-tailed 
distributions have a chance to be more appropriate.\\ 
Why? Let $X_1$, $X_2$ be insurance claims due to flood disasters $(X_1)$ and wind storms $(X_2)$. 
Last year events taught us that very often the extreme values of $X_1$ are accompanied by extreme values of $X_2$. 
In mathematical language it means, that$X_1$ and $X_2$ are asymptotically dependent. Traditional models based 
on multidimensional normal probability law give rise to quite opposite conclusion. Therefore in modelling 
of extreme events more and more often the researchers use the wider class of distributions, which includes 
the normal distribution as a special case. \\    
Let us now consider a simple example from finance market. On the figure below you see daily log-returns 
of stock indices DAX and CAC ( horizontal axis $X_1$-DAX, vertical axis $X_2$-CAC). The data cover the period 
from 1990 to 2004 (about 4000 data). The scatter plot assumes a shape of "elliptic cloud". And the level sets of 
the probability density, of the random vector $(X_1, X_2)$, are ellipses. This empirical observation suggests, 
that the family of elliptic distributions should be taken under consideration.\\
\begin{figure}
\includegraphics*[width=12cm,height=7cm]{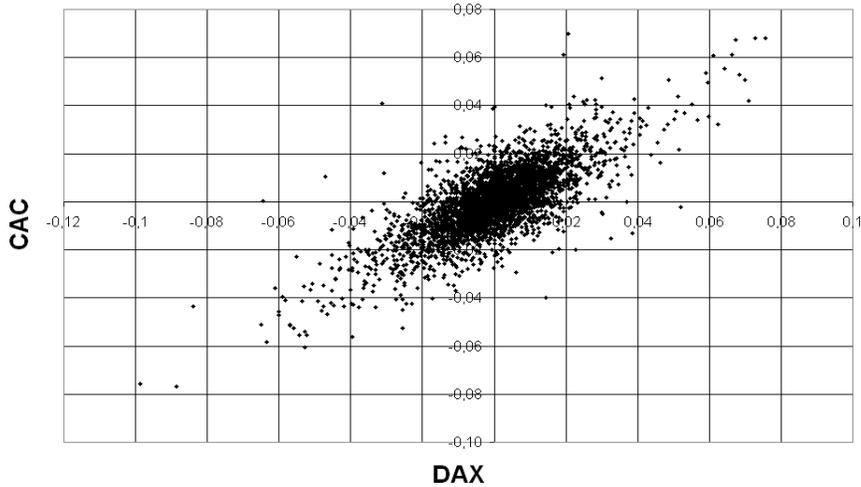}
\caption{Log-returns of DAX and CAC.}
\end{figure}
Furthermore we can ask  how often we observe the situation, when the daily log-returns of the both indices take
 the extreme values.\\
Let $W(j)$ be the quantity of observations $(x_{1,k},x_{2,k})$ such that \\$(x_{1,k}> x_{1,j}, x_{2,k}>x_{2,j})$,
 where $x_{i,j}$ is the j-th order statistics of the random variable $X_i$, $i=1,2$.\\
\begin{figure}
\includegraphics*[width=12cm,height=7cm]{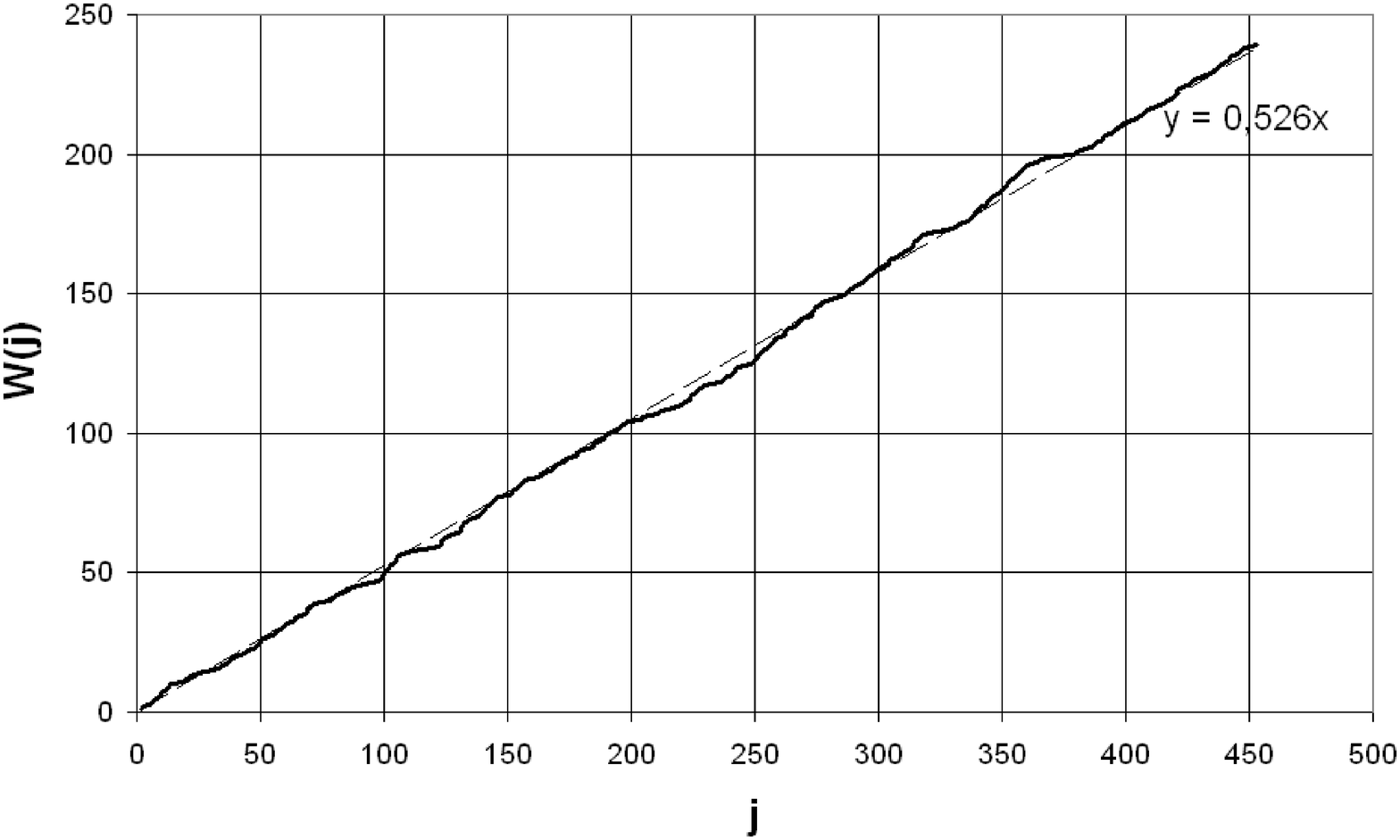}
\caption{Graph of $W(j)$.}
\end{figure}

The graph of the function $W(j)$ shows us, that $X_1$ and $X_2$ are asymptotically dependent. So the joint 
random variable $(X_1, X_2)$ couldn't be normally distributed.\\

\section{Preliminaries}
To begin with, we recall the basic definitions.\\
Let $F$ be an univariate distribution function and $F^{-1}$ its generalized inverse \\
\[F^{-1}(u)=\inf \{ x\in \mathbb{R} \: :\: F(x)\geq u \} \mbox{ for all } u \in (0,1).\]

\begin{definition}
Let $(X_1, X_2)$ be a random vector with marginal distribution functions $F_1$ and $F_2$.\\
The coefficient of upper tail dependence of $(X_1,X_2)$ is defined to be\\
\[\lambda_U (X_1, X_2) = \lim_{u\rightarrow 1} P(X_1>F_1^{-1}(u) | X_2>F_2^{-1}(u))\;\;\;,\] 
provided, that the limit $\lambda_U \in [0,1]$ exists.\\
If $\lambda_U=0$ , then we say that $X_1$ and $X_2$ are asymptotically independent.\\
Otherwise ( that is $\lambda_U >0$ or $\lambda_U$ doesn't exists ) we say that, they are asymptotically dependent.\\
\end{definition}

For a pair of random variables upper tail dependence is a measure of joint extremes. That is 
they measure the probability that one component is at an extreme of size given that the other is at the same
extreme, relative to the marginal distributions.\\

\begin{lemma}
If two continuously distributed random variables $X_1$, $X_2$ are independent, then they are asymptotically
 independent.\\
Proof.\\
\[\lim_{u\rightarrow 1} P(X_1>F_1^{-1}(u) | X_2>F_2^{-1}(u))= \lim_{u\rightarrow 1} P(X_1>F_1^{-1}(u))=0\;\;\;.\]
\end{lemma}
Note that the bivariate normal distribution has the same property.\\

The tail behaviour can be also described in a "symmetric way".\\

\begin{definition}
The bivariate random variable $X=(X_1,X_2)$ is said to be regularly varying with index $\beta >0$, if for all $y>0$
 and for every angle $[\alpha_0, \alpha_1]$
\[\lim_{t\rightarrow \infty} \frac{P(|X|>ty, \arg X\in[\alpha_0,\alpha_1])}{P(|X|>t)}=y^{-\beta}M(\alpha_0,\alpha_1).\]
where\\
$M(\alpha_0,\alpha_1)$ is a certain measure on the interval $[0,2\pi)$.
\end{definition}

\begin{definition}
If $X$ is a bivariate random variable and, for some \\
$\mu \in \mathbb{R}^2$, some 2x2 nonnegative definite symmetric
matrix $\Sigma$ and some function $\psi\: :\:[0,\infty) \rightarrow \mathbb{R}$, the characteristic function is of 
the form \[\varphi(t)=exp(it^* \mu) \psi(t^* \Sigma t)\;\;\;,\]
 then we say that $X$ has an elliptical distribution with parameters
 $\mu, \Sigma$ and $\psi$, and we write $X\sim E_2(\mu, \Sigma, \psi)$.  
\end{definition}

The function $\psi$ is referred to as the characteristic generator of $X$.\\

\begin{remark}
The following widespread used distributions prove to be\\ elliptic:\\
1. the normal distribution\\
   \[\psi(t^2)=exp(\frac{-t^2}{2})\;\;\;,\]   
2. some $\alpha$-stable with characteristic generator of the form\\
   \[\psi(t^2)=exp(\frac{-|t|^{\alpha}}{2})\;\;\;,0< \alpha <2\;\;\;,\]
3. T-Student distribution.\\
\end{remark}

\section{Result}
\begin{theorem}
Let
\[\phi(t)=\psi(t^* \Sigma t)\;\;\;,t\in \mathbb{R}^2\]
be a characteristic function of a bivariate elliptically distributed random variable $X=(X_1,X_2)$.\\
1. $\Sigma$ is a positive definite symmetric matrix,\\
2. $\psi:\mathbb{R_+}\longrightarrow \mathbb{R}$ is such that:\\
\hspace*{2em}   $\psi(r^2)=\psi_0(r^2)+r^{\beta}\psi_1(r^{\gamma})$,  $0< \gamma \leq 2$,\\
\hspace*{2em}   $\beta\in\mathbb{R_+} \setminus 2\mathbb{N},\;\; $\\
\hspace*{2em}   $\psi_0 ,\psi_1\in\mathbb{C^{\infty}}(\mathbb{R_+})\;\; \wedge\;\; \psi_1(0)\neq0
\;\; \wedge\;\; \psi_0(0)=1$,\\
\hspace*{2em}   $\forall\;0\leq k \leq 4+[\beta]\;\;\;$ $\lim_{t \rightarrow \infty}t^{k+\frac{1}{2}}\psi^{(k)}(t^2)=0 $,\\
then the marginal random variables $X_1$ and $X_2$ are asymptotically dependent .\\
\end {theorem}

\section{Concluding remarks}
\noindent
1. For all $\alpha \in (0,2)$ , if $X=(X_1, X_2)$ is elliptic and $\alpha$- stable, then $X_1$ and $X_2$ are
   asymptotically dependent.\\
2. The result is also valid for the characteristic generator of the form\\
$\psi(r^2)=\psi_0(r^2)+r^{\beta_1}\psi_1(r^{\gamma_1})+ ... +r^{\beta_m}\psi_m(r^{\gamma_m})$, where\\
$ 0< \beta_1 < \beta_2 <...< \beta_m $.\\

\section{Proof of the Theorem}
\begin{lemma}
Let us have the same assumptions as in theorem 3.1.\\
Then the asymptotics of the probability density of bivariate random variable\\
 $X=(X_1, X_2)$ formulates as follows 
\[g(x)= c ||x||^{-2-\beta} + O( ||x||^{-3-\beta})\;\;,\;\;||x||\rightarrow \infty\;\;,\;\;c=const>0\;\;,\]
$||x||=\sqrt{x^* \Sigma x}$  ,  $x=(x_1, x_2)$.\\
\end{lemma}  
Proof.\\

\[\lambda(u)=P(X_1 > F^{-1}_1(u)|X_2 > F^{-1}_2(u))=\frac{P(X_1 > F^{-1}_1(u) \wedge X_2 > F^{-1}_2(u))}
{P(X_2 > F^{-1}_2(u))}\]
Let us denote  $F^{-1}_j(u)=as_j,\;\;j=1,2;\;\;s_1,s_2=const>0,\;\;a>>0$\\
\[P(X_1 > F^{-1}_1(u) \wedge X_2 > F^{-1}_2(u))=P(X_1 > as_1 \wedge X_2 > as_2)=\]
\[=
\int_{as_1}^{+\infty}dx_1 \int_{as_2}^{+\infty}dx_2 \int_{\mathbb{R}^2}e^{-ix^* t} \psi(t^*\Sigma t) dt\]

We calculate the asymptotics of the integral above, for $a\longrightarrow +\infty$.\\
$\Sigma>0$ and symmetric $\Longrightarrow\;\;\Sigma=A^*A$.\\ 
\[g(x)=\int_{\mathbb{R}^2}e^{-ix^* t} \psi(t^*\Sigma t) dt=
\int_{\mathbb{R}^2}e^{-ix^*A^{-1}w} \psi(w^*w)(det\Sigma)^{-\frac{-1}{2}} dw\;\;,\]
after the change of the variables $t=A^{-1}w$.\\
Next we substitute $x=A^*y$ and obtain \\
\[g(A^*y)=\int_{\mathbb{R}^2}e^{-iy^*w} \psi(w^*w)(det\Sigma)^{-\frac{-1}{2}} dw=(det\Sigma)^{\frac{-1}{2}}G(y)\;\;.\]
We change the variables a second time $w_1=r\cos\varphi , w_2=r\sin\varphi$  ,
 and let us express $y_1 , y_2$ in the form 
$y_1=||y||\sin\alpha , y_2=||y||\cos\alpha$.\\
 Then
\[G(y)=\int_0^{+\infty}r\psi(r^2)dr \int_0^{2\pi}e^{-ir||y||\sin(\varphi+\alpha)}d\varphi=\]
\[=
\int_0^{+\infty}r\psi(r^2)dr \int_{\alpha}^{2\pi+\alpha}e^{-ir||y||\sin\varphi}d\varphi=
2\pi\int_0^{+\infty}r\psi(r^2)J_0(r||y||)dr\;\;,\]
where $J_0$ is Bessel function.\\
\[J_0(r||y||)=\frac{1}{\sqrt{\pi r||y||}}[\cos(r||y||-\frac{\pi}{4})+O((r||y||)^{-1})]\;\;,\]
for $r||y||\longrightarrow \infty$ i $|\arg  r||y|| |\leq\pi-\epsilon<\pi$ , cf.\cite{[F]}\\

\[\int_0^{+\infty}\frac{r\psi(r^2)}{\sqrt{\pi r||y||}}\cos(r||y||-\frac{\pi}{4})dr=
\frac{1}{\sqrt{\pi||y||}}\;Re\; e^{\frac{-i\pi}{4}}\int_0^{+\infty}e^{ir||y||}r^{\frac{1}{2}}\psi(r^2)dr\;\;.\]
Now we compute the first term of asymptotics of the integral \\
\[F(||y||)=\int_0^{+\infty}e^{ir||y||}r^{\frac{1}{2}}\psi(r^2)dr\;\;\;,for \;\;||y||\longrightarrow  \infty\;\;.\]
We assumed that the function $\psi(r^2)$ and its derivatives tend quickly to $0$, as $r \rightarrow \infty$. 
 Therefore with the help of localization rule and Erdelyi Lemma cf.\cite{[F]} we obtain\\
\[a)\;\;Re[ e^{\frac{-i\pi}{4}}\int_0^be^{ir||y||}r^{\frac{1}{2}}\psi_0(r^2)dr]=0\;\;,\;\;
\mbox{ the asymptotics is trivial, }\]
\[b)\;\;e^{\frac{-i\pi}{4}}\int_0^be^{ir||y||}r^{\frac{1}{2}+\beta}\psi_1(r^{\gamma})dr=\]
\[=
 e^{\frac{-i\pi}{4}}\psi_1(0)\Gamma(\frac{3}{2}+\beta)e^{\frac{i\pi(\frac{3}{2}+\beta)}{2}}||y||^{\frac{-3}{2}-\beta}+
+O(||y||^{\frac{-5}{2}-\beta})=\]
\[=ie^{\frac{i\pi\beta}{2}}
\psi_1(0)\Gamma(\frac{3}{2}+\beta)e^{\frac{i\pi(\frac{3}{2}+\beta)}{2}}||y||^{\frac{-3}{2}-\beta}+
O(||y||^{\frac{-5}{2}-\beta})\;\;\;,\]
\[Re [ie^{\frac{i\pi\beta}{2}}\psi_1(0)
\Gamma(\frac{3}{2}+\beta)e^{\frac{i\pi(\frac{3}{2}+\beta)}{2}}||y||^{\frac{-3}{2}-\beta}]=
-\sin{\frac{\pi\beta}{2}}\psi_1(0)\Gamma(\frac{3}{2}+\beta)||y||^{\frac{-3}{2}-\beta}\]
The expression above isn't trivial, when $\beta$ is not a natural even number.\\
Hence the first term of the asymptotics of the integral $G(y)$ is given by a formula:\\
\[G(y)=2\sqrt{\pi}(-\sin{\frac{\pi\beta}{2}})\psi_1(0)\Gamma(\frac{3}{2}+\beta)||y||^{-2-\beta}+O(||y||^{-3-\beta})=\]
\[
=c_1||y||^{-2-\beta}+O(||y||^{-3-\beta})\;\;\;for \;\;||y||\longrightarrow  \infty\;\;.\]
Thus\\
\[g(x)=c_1(det\Sigma)^{\frac{-1}{2}}||(A^{-1})^*||^{-2-\beta} ||x||^{-2-\beta}+O(||x||^{-3-\beta})\;\;.\]

\begin{lemma}
Under the assumptions of theorem 3.1 the bivariate random variable $X=(X_1, X_2)$ is regularly varying with the index 
$\beta$.
\end{lemma}
Proof.\\
\[ \frac{P(|X|>ty , \arg X\in [\alpha_0, \alpha_1])}{P(|X|>t)} =
 \frac{\int_{ty}^{+\infty}\;\;\int_{\alpha_0}^{\alpha_1}\;\;r g(r,\theta)\;d\theta\;dr}
{\int_t^{+\infty}\;\;\int_0^{2\pi}\;rg(r,\theta)\;d\theta\;dr}=\]
\[=
\frac{(\alpha_1-\alpha_0) \int_{ty}^{+\infty}\;\;[r^{-1-\beta}+ O(r^{-2-\beta})]\;dr}
{2\pi \int_t^{+\infty}\;\;[r^{-1- \beta}+ O(r^{-2- \beta})]\;dr}=
\frac{\alpha_1- \alpha_0}{2\pi}\;y^{-\beta}\;\;\;.\] 
Lemma 5.2 implies the thesis of Theorem 3.1 cf.\cite{[HL]}.\\

\pagebreak

\end{document}